\documentclass[journal=jacsat,manuscript=article]{achemso}

\usepackage{chemformula}
\usepackage[T1]{fontenc}
\usepackage{comment}
\usepackage{algorithm}
\usepackage{rotating}
\usepackage[noend]{algpseudocode}
\usepackage{tikz}
\usetikzlibrary{shapes, arrows}
\usepackage{caption}
\usepackage{subcaption}
\usepackage[font=small]{caption}
\usepackage[labelformat=empty, position=top]{subcaption}
\usepackage{cancel}
\usepackage{changepage}
\usepackage{amsfonts}
\usepackage{amsmath}
\usepackage{mathtools}
\usepackage{ulem}
\usepackage{bigints}

\usepackage{modiagram_edit}
\usepackage{hyperref}
\usepackage{graphicx}
\usepackage{caption}
\usepackage{kbordermatrix}% http://www.hss.caltech.edu/~kcb/TeX/kbordermatrix.sty
\usepackage{braket}

\graphicspath{ {./} }

\title{Benchmarking DFT-based excited-state methods for intermolecular charge-transfer excitations}

\author{Nicola Bogo}
\affiliation[Uni-DuE]
{Faculty of Physics, University of Duisburg-Essen, 47057 Duisburg, Germany}

\alsoaffiliation[TUM]
{Department of Chemistry and Catalysis Research Center, TUM School of Natural Sciences, Technische Universität München, Lichtenbergstr. 4, 85748 Garching, Germany}

\author{Christopher J. Stein}
\email{christopher.stein@tum.de}

\affiliation[TUM]
{Department of Chemistry and Catalysis Research Center, TUM School of Natural Sciences, Technische Universität München, Lichtenbergstr. 4, 85748 Garching, Germany}
\begin{document}

\maketitle

\begin{abstract}
    Intermolecular charge-transfer is a highly important process in biology and  energy-conversion applications where generated charges need to be transported over several moieties. However, Its theoretical description is challenging since the high accuracy required to describe these excited states must be accessible for calculations on large molecular systems. In this benchmark study, we identify reliable low-scaling computational methods for this task. Our reference results were obtained from highly accurate wavefunction calculations that restrict the size of the benchmark systems. However, the density-functional theory based methods that we identify as accurate can be applied to much larger systems. Since targeting charge-transfer states requires the unambiguous classification of an excited state, we first analyze several charge-transfer descriptors for their reliability concerning intermolecular charge-transfer and single out $D_\text{CT}$ as an optimal choice for our purposes. In general, best results are obtained for orbital-optimized methods --- and among those, IMOM proved to be the most numerically stable variant --- but optimally-tuned range-separated hybrid functionals combined with rather small basis sets proved to yield surprisingly good results. This makes these fast calculations attractive for high-throughput screening applications.
\end{abstract}

\section{Introduction}

Charge-transfer (CT) excitations between two molecules where one molecular unit serves as the donor and the other as the acceptor of the transferred electron are classified as intermolecular charge transfer (ICT) if these units are individual molecules or moieties sufficiently far apart.
 Biological molecular devices exploit these transfer properties to move excess energy in the excited state through different locations in space and perform complex operations\cite{vinyard2013photosystem, sirohiwal2023reaction}. Naturally, chemists strive to understand the properties of such molecular systems to engineer novel, artificial devices with accurately predicted CT properties.
 
In supramolecular chemistry, large multi-component systems are synthesized starting from separate molecular units. The key benefit of the assembled structure is that the characteristics of the isolated components are preserved, whereas novel properties emerge that arise due to the arrangement of the individual units
in the nanostructure.
Supramolecular cage molecules hence provide a blueprint for such artificial CT devices and since donor and acceptor moieties are well separated due to the presence of linkers, spacers, and connecting metal ions, we focus our investigation on molecular dimers with a donor and acceptor unit as prototypes for the study of ICT processes. Unfortunately, accurate wavefunction methods are plagued by a prohibitively steep scaling of the computational cost, and the typical size of supramolecular cages forces us to apply more economic electronic-structure methods. Hence, in this work, we benchmark DFT-based methods to identify the most accurate methods suitable for calculations on systems with more than a hundred atoms. Specifically, we apply time-dependent DFT (TD-DFT) within the Tamm--Dancoff approximation (TDA), and orbital-optimized DFT methods (OO-DFT), and benchmark them against highly accurate wavefunction theory on small dimer systems to investigate their accuracy for ICT excitations.

In the following, we first review a selection of descriptors for CT states and several DFT-based excited-state methods. We then discuss the strengths and weaknesses of these methods in describing ICT for molecular dimers at different distances.

\section{Charge-Transfer descriptors}

In organic photochemistry, excitations are frequently divided into 4 classes: valence excitations, core excitations, Rydberg excitations, and charge-transfer excitations. The identification of the CT character of a given excited state is a crucial prerequisite for any study of CT properties. Many approaches were proposed for this problem, resulting in a diverse set of CT descriptors. Here, we will divide them into two major classes: density-based and wavefunction-based descriptors. Appropriate descriptors are expected to yield a consistent classification, while quantitative results may differ. Obviously, only descriptors with the same unit and range can be directly compared quantitatively.

    \textbf{Fragment localized-orbital-based desriptors}\\
    Plasser \textit{et al.} defined two sets of CT descriptors, both derived from the analysis of the CIS-like wavefunction produced by a TD-DFT calculation under the TDA and the resulting 1-particle transition-density matrix (1TDM). The first set\cite{plasser2012analysis} originates from the construction of an approximate 1TDM in a two-fragment model for each excited state.
    
    In a basis of localized molecular orbitals, all orbitals can be assigned to a predefined fragment A or B. For the sake of simplicity, we consider a molecular dimer system consisting of two separate molecular units which constitute the two fragments.
    Excited states dominated by a singly-excited configuration can be produced from the ground state by applying a spin-averaged excitation operator\cite{east2000naphthalene} $\hat{E}_{sr}$ on the ground state $\ket{0}$

    \pagebreak 
    
    \begin{center}
     \vspace{15pt}
        
        \begin{modiagram}
          \AO[a]{s}{0;pair}         \AO[b]{s}{1;}
          \AO[c](1cm){s}{0;pair}       \AO[d](1cm){s}{1;}
          \AO[e](3cm){s}{0;down}    \AO[f](3cm){s}{1;}
          \AO[g](4cm){s}{0;pair}    \AO[h](4cm){s}{1;up}
          \draw[red,->] (a.-0) -- (d.180) node {};
          \draw[->] (1.7cm,0.5cm) -- (2.2cm,0.5cm) node {};
          \node[xshift=6.7cm,yshift=0.5cm]{$=\ket{\text{A}^+\text{B}^-}=\frac{1}{\sqrt{2}} \ \hat{E}_{f'i} \ket{0}$};
        \end{modiagram}
        
        \vspace{30pt}
        
        \begin{modiagram}
          \AO[a]{s}{0;pair}\node[yshift=-1cm]{ $\mathrm{A}$};   \AO[b]{s}{1;}
          \AO[c](1cm){s}{0;pair}\node[xshift=1cm,yshift=-1cm]{ $\mathrm{B}$};   \AO[d](1cm){s}{1;}
           \AO[e](3cm){s}{0;pair}\node[xshift=3cm,yshift=-1cm]{ $\mathrm{A}$};   \AO[f](3cm){s}{1;up}
          \AO[g](4cm){s}{0;down}\node[xshift=4cm,yshift=-1cm]{ $\mathrm{B}$};    \AO[h](4cm){s}{1;}
          \draw[red,->] (c.180) -- (b.0) node {};
          \draw[->] (1.7cm,0.5cm) -- (2.2cm,0.5cm) node {};
          \node[xshift=6.7cm,yshift=0.5cm]{$=\ket{\text{A}^-\text{B}^+}=\frac{1}{\sqrt{2}} \ \hat{E}_{fi'} \ket{0}$};
        \end{modiagram}
        
    \end{center}
    
    where $r=i, i'$ and $s= f, f'$ and only the ionic contributions $\ket{\text{A}^+}$, $\ket{\text{A}^-}$, $\ket{\text{B}^+}$ and $\ket{\text{B}^-}$ arising from a CT transition are shown. The $\hat{E}_{sr}$ operators are \textit{spin-averaged} excitation operators and the normalization $\frac{1}{\sqrt{2}}$ comes from the spin averaging (see Ref.\citenum{east2000naphthalene}).
    
    The 1TDM $\mathbf{D^{0\alpha}}$ for each excited state $\ket{\alpha}$ can be written in the localized basis $\{i, f, i', f'\}$ of MOs centered on fragment A $\{i, f\}$ and B $\{i', f'\}$, where transitions occur from row to column. Each element of $\mathbf{D^{0\alpha}}$ is the amplitude of a spin-averaged \textit{de-excitation} operator $\hat{E}_{rs}$:
    \begin{equation}
    D^{0\alpha}_{rs} = \bra{0} \hat{E}_{rs} \ket{\alpha}
    \end{equation}
    
    since $\hat{E}_{rs}$ acts on excited state $\alpha$. \textit{E.g.} for the CT configuration
    
    \begin{center}
        $\ket{\beta}=\ket{\text{A}^-\text{B}^+}=\frac{1}{\sqrt{2}} \ \hat{E}_{fi'} \ket{0}$
    \end{center}
    we find
    \renewcommand{\kbldelim}{(}% Left delimiter
    \renewcommand{\kbrdelim}{)}% Right delimiter
    \[
      \mathbf{D^{0\beta}} = \kbordermatrix{
        & i & f & i' & f'  \\
        i & 0 & 0 & 0 & 0 \\
        f & 0 & 0 & 0 & 0 \\
        i' & 0 & \sqrt{2} & 0 & 0 \\
        f' & 0 & 0 & 0 & 0
      }
    \]\,

    With unprimed orbitals located on A and primed orbitals located on B, a block structure emerges for  $\mathbf{D^{0\alpha}}$
    $$
    \mathbf{D^{0\alpha}}=\left( 
    \begin{array}{c|c} 
      \mathrm{A} \xrightarrow{} \mathrm{A} & \mathrm{A} \xrightarrow{} \mathrm{B} \\ 
      \hline 
      \mathrm{B} \xrightarrow{} \mathrm{A} & \mathrm{B} \xrightarrow{} \mathrm{B}
    \end{array} 
    \right) 
    $$
    
    Diagonal blocks $\mathrm{A} \xrightarrow{} \mathrm{A}$ and $\mathrm{B} \xrightarrow{} \mathrm{B}$ pertain to local excitations contributing to the CIS wavefunction, while off-diagonal blocks $\mathrm{A} \xrightarrow{} \mathrm{B}$ and $\mathrm{B} \xrightarrow{} \mathrm{A}$ contain contributions of charge-transfer excitations involving both fragments.\\
    Once the 1TDM for an excited state $\alpha$ has been constructed under a given partitioning scheme, a 2-by-2 charge-transfer number matrix $\mathbf{\Omega^\alpha}$ can be constructed by summing over its squared elements. Please note that for the sake of simplicity, we assume that all localized orbitals can be unambiguously assigned to fragment A or B. An element of $\mathbf{\Omega^\alpha}$ is hence defined as
    \begin{equation}
        \Omega_\mathrm{AB}^\alpha=\frac{1}{2} \sum_\mathrm{a \in A} \sum_\mathrm{b \in B} ( D_{ab}^{\mathrm{0} \alpha} )^{2}.
    \end{equation}
    The off-diagonal elements of this matrix then quantify the weight of CT excitations in the excited state\cite{luzanov2010electron}. 
    
    \vspace{0.5cm}
    
    The elements of $\mathbf{\Omega}$ are then used to compute CT descriptors characterizing the locality of the excitation from the ground state to the target excited state. To make the notation more compact, we will drop the $\alpha$ superscript in the definition of the CT descriptors from here on, as it is obvious that each excited state needs to be analyzed separately. Based on this, the charge transfer descriptor $CT$ is defined by simply summing over the off-diagonal elements of $\mathbf{\Omega}$
    \begin{equation}
        CT={\frac{1}{N}} \sum_\mathrm{A} \sum_\mathrm{B \neq A}\Omega_{\mathrm{AB}}\, ,
    \end{equation}
    where $N$ is the norm of the 1TDM, approximately 1 for single excitations. $CT$ is 0 when there are no off-diagonal elements, while $CT=1$ if the CIS wavefunction involves only contributions from the excitations between fragment A and fragment B. By summing over the elements of $\mathbf{\Omega}$, Le Bahers \textit{et al.} define many other descriptors.\cite{plasser2012analysis} 
    
    Two limitations apply to this approach. One is related to the fact that the 1TDM analysis works with a CIS-like wavefunction; this is the form characterizing the results of linear-response (LR) TD-DFT, but higher-order response TD-DFT includes contributions from higher-order excitations to some extent. 
    Another limitation arises for the application of the 1TDM analysis to methods that approximate excited states with a single configuration and subsequently relax the density of such a configuration. This is the case for the OO-DFT methods we apply in this study. In these methods, since the MOs for the ground and excited states are different, it is not trivial to apply the 1TDM analysis.
    Moreover, neither the partitioning of a molecular system into two fragments nor the orbital localization is unique. This is particularly problematic when the excited-state analysis is applied to single chromophores, while it is admittedly rather trivial when the CT-numbers matrix is computed for a molecular dimer.
    
     \textbf{Electron-hole distance descriptor}\\
    The second set\cite{plasser2015statistical} of parameters from Plasser \textit{et al.} is focused on the interpretation of the 1TDM as a 2-particle excitonic wavefunction. In many-body Green's function theory, the 1TDM is a 2-body wavefunction $\chi$, describing the correlated motion of \textit{hole} and \textit{electron} \textit{quasi}-particles in the excited state $\alpha$
    \begin{equation}
        \chi_{\mathrm{exc}}(\textbf{r}_{\mathrm{h}},\textbf{r}_{\mathrm{e}})=N\int\Phi^{0}(\textbf{r}_{h},\textbf{r}_{2},\ldots,\textbf{r}_{N})\times\Phi^{\alpha}(\textbf{r}_{e},\textbf{r}_{2},\ldots,\mathbf{r}_{N})\ \mathrm{d}\mathbf{r}_{2}\ldots\ \mathrm{d}\mathbf{r}_{N}
    \end{equation}
    where $\Phi^{0}$ and $\Phi^{\alpha}$ are the ground and excited state $\alpha$ wavefunctions respectively, and $\textbf{r}_i$ denotes the coordinates of the $i$-th electron. The matrix representation of the 1TDM enables the expansion of $\chi_{\mathrm{exc}}$
        
    \begin{equation}
    D_{\mu\nu}^{0\alpha}=\langle\Phi^{0}|\hat{a}_{\mu}^{\dagger} \hat{a}_{\nu}|\Phi^{\alpha}\rangle \quad \xrightarrow{} \quad \chi_{\mathrm{exc}}(\textbf{r}_{\mathrm{h}},\textbf{r}_{\mathrm{e}})=\sum_{\mu\nu}\  D_{\mu\nu}^{0\alpha}\ \chi_{\mu} (\textbf{r}_{\mathrm{h}})\ \chi_{\nu}(\textbf{r}_{\mathrm{e}})\, .
    \end{equation}
    Since $\chi_{\mathrm{exc}}$ is a two-particle wavefunction, its square $\chi_{\mathrm{exc}}^2$ computed for given $\textbf{r}_{\mathrm{e}}$ and $\textbf{r}_{\mathrm{h}}$ returns the probability of having the electron in position $\textbf{r}_{\mathrm{e}}$, while the hole is in position $\textbf{r}_{\mathrm{h}}$. $\chi_{\mu}(\textbf{r}_{\mathrm{h}})^2$ and $\chi_{\nu}(\textbf{r}_{\mathrm{e}})^2$ can be studied separately like the density-accumulation and depletion functions by Le Bahers \textit{et al.} covered in the next paragraph, providing similar information as for their density-based parameters. \\
    The advantage of $\chi_{\mathrm{exc}}$ depending explicitly on the electron and hole positions is that parameters characterizing the electron-hole correlation can be computed. One example is the calculation of the $d_{\mathrm{exc}}$ parameter\cite{mewes2016excitons}, also termed $\mathrm{RMS}d_{eh}$ in Ref.~\citenum{wang2023earth} or RMSeh in the documentation of the TheoDORE package\cite{plasser2020theodore}:
    \begin{equation}
        d_{\mathrm{exc}}={\sqrt{\langle|{\textbf{r}}_{e}-{\textbf{r}}_{h}|^{2}\rangle}}
    \end{equation}
    which measures the root-mean-square distance in space between the hole and the electron, assessing the size of the exciton.
        
    \textbf{Density-based descriptors}\\
    Le Bahers, Adamo, and Ciofini \cite{le2011qualitative} provide an alternative characterization of the changes in the electron density happening upon excitation.
    The ground-state electron density $\rho_{\mathrm{GS}}(\textbf{r})$ is subtracted from the excited-state electron density $\rho_{\mathrm{EX}}(\textbf{r})$ producing the density-difference function $\Delta\rho(\textbf{r})$, and its positive and negative co-domains $\rho_{+}(\textbf{r})$ and $\rho_{-}(\textbf{r})$ are characterized separately. 

    \begin{equation}
    \Delta\rho(\textbf{r})=\rho_{\mathrm{EX}}(\textbf{r})-\rho_{\mathrm{GS}}(\textbf{r})
    \end{equation}    
    %\end{itemize}
    %
    \begin{equation}
        \rho_{+}(\textbf{r})\,=\,\left\{\begin{array}{c l}{{{\Delta\rho}(\textbf{r})}}&{{\mathrm{if}\;\Delta\rho(\textbf{r})> 0}}\\ {{0}}&{{\mathrm{if}\;\Delta\rho(\textbf{r})< 0}}\end{array}\right. \quad \rho_{-}(\textbf{r})\,=\,\left\{\begin{array}{c l}{{{\Delta\rho}(\textbf{r})}}&{{\mathrm{if}\;\Delta\rho(\textbf{r})< 0}}\\ {{0}}&{{\mathrm{if}\;\Delta\rho(\textbf{r})> 0.}}\end{array}\right.
    \label{+-densities}
    \end{equation}

    \vspace{0.5cm}
    In areas where the density-difference function is positive, charge accumulates upon excitation, whereas in areas where the density-difference function is negative, charge is depleted upon excitation. Each density-accumulation and depletion distribution can be characterized by studying its center of charge
    \begin{equation}
        \textbf{R}_{+/-}=\frac{\displaystyle\int \textbf{r}\ \rho_{+/-}(\textbf{r})\mathrm{d}\textbf{r}}{\displaystyle\int\rho_{+/-}(\textbf{r})\mathrm{d}\textbf{r}},
    \end{equation}
    and variance along the \textit{x, y} and \textit{z} directions
    \begin{equation}
        \sigma_{a,j}={\sqrt{\frac{\sum_{i}\rho_{a}(\textbf{r}_{i})(j_{i}-j_{a})^{2}}{\sum_{i}\rho_{a}(\textbf{r}_{i})}}} \quad \mathrm{where} \quad j=x,y,z; \, \,a=+\ \mathrm{or}\ - .
    \end{equation}
    The modulus of the distance between the centers of charge accumulation and depletion $\textbf{R}_+$ and $\textbf{R}_-$ defines the $D_{\mathrm{CT}}$ descriptor, in units of distance (\r{A} in this work). $\mu_{\mathrm{CT}}$ is the dipole moment obtained by multiplying the amount of displaced charge $q_{\mathrm{CT}}$ times the $D_{\mathrm{CT}}$ parameter, in units of charge times distance. For TD-DFT calculations, these two parameters always return the same number since $q_{\mathrm{CT}} = 1$. This is not always true for OO-DFT methods, as the positive and negative co-domains of the electron density-difference distributions frequently do not integrate to one electron ($q_{\mathrm{CT}} \neq 1$).
    \begin{equation}
        D_{\mathrm{CT}}=|\textbf{R}_{+}-\textbf{R}_{-}|, \quad \left|\left|\mu_{\mathrm{CT}}\right|\right|=D_{\mathrm{CT}}\int\rho_{+}(\textbf{r})\ \mathrm{d}\textbf{r}=D_{\mathrm{CT}}\ q_{\mathrm{CT}} \, .
    \end{equation}
    Ellipsoid functions can be defined at the centers of charge accumulation and depletion ($\textbf{R}_+$ and $\textbf{R}_-$) with axis length equal to the variance of the electron density-accumulation/depletion distributions along each direction,
    \begin{equation}
        C_{a}(\textbf{r})=A_{a}\ \mathrm{exp}\!\left(-\frac{(x-x_{a})^{2}}{2\sigma_{ax}^{2}}-\frac{(y-y_{a})^{2}}{2\sigma_{ay}^{2}}-\frac{(z-z_{a})^{2}}{2\sigma_{az}^{2}}\right) \quad \mathrm{where} \quad a=+\,\mathrm{or}\,-
    \end{equation}
    and $A_a$ is a normalization factor. The product of normalized $C_+$ and $C_-$ can be integrated, producing $S_{\mathrm{+-}}$, as proposed by Lu and implemented in the Multiwfn\cite{lu2012multiwfn} program
    \begin{equation}
       %notation from Multiwfn manual
       %S_{+-}=\bigintsss\sqrt{\frac{C_{+}({r})}{A_{+}}}\sqrt{\frac{C_{-}({r})}{A_{-}}}\ \mathrm{d}{r}
       S_{+-}=\int\sqrt{C_{+}({\textbf{r}})}\sqrt{C_{-}({\textbf{r}})}\ \mathrm{d}{\textbf{r}}\, .
       %since it's normalized, we can write it like this:
       %S_{+-}=\int C_{+}({r})\ C_{-}({r})\ \mathrm{d}{r} \, .
    \end{equation}
    %
    %where $A_{+}$ and $A_{-}$ are normalization factors, neglected for our definition of $C_{+}$ and $C_{-}$. 
    This dimensionless descriptor $S_{+-}$ is 1 for local excitations and drops to 0 by displacing the $C_{+}$ and $C_{-}$ far from one another in an ICT state. \\
    The obvious advantage of applying density-based CT quantifiers is that they can be computed using normalized electron density cube files, easily produced by commercially available electronic-structure codes. We would like to point out, however, that parameters based on the center of charge like $D_{\mathrm{CT}}$ are prone to fail for charge-resonance excitations.

    \textbf{Earth-mover distance}\\
    A novel density-based CT descriptor\cite{wang2023earth}, inspired by the solution of the \textit{Earth-Mover Distance} (EMD) optimal transport problem, has been introduced by Wang, Liang, and Head-Gordon. Interestingly, a similar CT descriptor was proposed around the same time in two other publications, respectively by Lieberherr, Gori-Giorgi and Giesbertz\cite{lieberherr2023optimal}, and Fraiponts, Maes and Champagne\cite{fraiponts2024earth}.

    The approach is based on the notion that changes in the electron density happening shortly after the excitation are similar to transport problems of discrete distributions. The EMD problem, extensively discussed in the field of computer vision, can be explained with the following analogy: assume a hole is dug in the ground, and the moved soil lies close by. The problem of transferring the soil back into the hole while minimizing the performed work is an optimal transport problem. The transportation simplex algorithm\cite{rubner1998metric} can be applied to solve it. By mapping the displacement of electron density happening after the excitation to an optimal transport problem, the charge-depletion and charge-accumulation distributions (equation \ref{+-densities}) appear as analogous to the hole ($\rho_-(\textbf{r})$) and soil ($\rho_+(\textbf{r})$) piles, denoted as the demand ($D$) and supply ($S$) piles. Operating with discrete density distributions computed on a three-dimensional grid, the density-difference distributions are characterized by point charges $q$ for a set of Cartesian coordinates $\textbf{r}$ of the grid points.
    \begin{equation}
        \rho_-(r) =\{(\mathbf{r}_{j},q_{j}^{D})\} = D \quad \mathrm{and} \quad \rho_+(r) =\{(\mathbf{r}_{i},q_{i}^{S})\} = S
    \end{equation}
    \begin{equation}
    {\sum_{i}q_{i}^{S}=\sum_{j}q_{j}^{D}}=q^{\mathrm{CT}}\, .
    \end{equation}
    In the optimal transport problem, the transmission matrix $\mathbf{F}$ is defined and its elements $f_{ij}$ are optimized such that
    \begin{equation}
        \mathbf{F}=\operatorname*{arg\ min}_{\mathbf{F}}\sum_{i j}f_{ij}d_{ij} \quad \mathrm{so\ that} \quad f_{ij} \ge 0, \quad \sum_{j}f_{ij}=q_{i}^{S} \quad \mathrm{and} \quad \sum_{i}f_{ij}=q_{j}^{D}\, .
    \end{equation}
    Finally, the charge-transfer descriptors $\mu^{\mathrm{EMD}}$ and $d^{\mathrm{EMD}}$ are defined based on that matrix
    \begin{equation}
        \mu^{\mathrm{EMD}}=\operatorname*{min}_{\mathbf{F}} \sum_{ij}f_{ij}d_{ij} \quad \mathrm{and} \quad d^{\mathrm{EMD}}={\frac{\mu^{\mathrm{EMD}}}{q^\mathrm{CT}}}
    \end{equation}
    where $\mu^{\mathrm{EMD}}$ has units of charge times distance and $d^{\mathrm{EMD}}$ has units of distance. Since it is based on an analysis of the density-difference distribution $\Delta\rho$, this parameter can be applied to any method capable of producing an approximated electron density distribution for the ground and excited state.
    
    Due to the scaling of the transportation simplex algorithm, one should employ a relatively coarse grid for the density analysis. Because of that, $\mu^\mathrm{EMD}$ and $\mathrm{d}^\mathrm{EMD}$ are computed on an Euler--MacLaurin grid.
    %is an Euler-MacLaurin grid always coarse? If not, just delete that sentence, we do not need it
    On the other hand, the density produced by a quantum-chemical calculation is interpolated on a quadrature grid, and the use of a thin mesh for the real-space grid is necessary to get accurate results. The calculation of EMD-based CT descriptors is available in the ChargeEMD open-source package for $\mu_{\mathrm{EMD}}$ and $d_{\mathrm{EMD}}$ on GitHub\cite{wang2023earth}.

More charge-transfer descriptors of various nature were proposed in the literature,\cite{richard2011time, peach2008excitation, guido2013metric, guido2014effective, etienne2014toward, ronca2014charge, ma2014electronic, etienne2015probing} but are not included in this work since some can be proven to be equivalent to the parameters introduced above.
\par In our study of ICT excitations, we will combine the descriptors that were introduced in the previous paragraph aiming to unravel the following information:

\begin{itemize}
  
   \item Find out if the computed excited state is a CT excitation, as this is not trivial for some methods like TD-DFT.
   %In TD-DFT, the excitation energy and excited-state properties are obtained by studying an approximation of the response function of the system and solving for its poles. This leads to the definition of an eigenvalue problem, which is solved iteratively and provides the excitation energies belonging to the requested \textit{n} lowest-energy excited states of the desired multiplicity, alongside an approximated CIS-like wavefunction belonging to each excited state. This can then be used to compute approximated transition dipole moments and an approximated electronic density for every requested excited state. 
   Frequently, when setting up a TD-DFT calculation the user does not know if a CT state is included in the requested lowest \textit{n} roots. If one or more CT states are present, it is not trivial to identify which roots show a strong CT character by simply looking at the dominant singly-excited configurations through visual inspection of the contributing orbitals.
  
   \item Classify the computed CT state as an inter-molecular CT state or an intra-molecular CT excitation centered on a single monomer.
    
   Excited states characterized by a moderate CT character centered on one single molecule (intra-molecular CT excitations) are standard in organic photochemistry. Their calculation is a very different problem compared to ICT, as the donor and acceptor MOs often overlap. This study focuses on ICT, and the best-performing methods for ICT are not necessarily the best-performing for intra-molecular CT. In the case of ICT, since the donor and acceptor molecular orbitals in the dominant configuration of a CT state are dislocated far in space, their overlap is expected to vanish; this overlap appears in the expression for the $K_{ab}$ term in Hartree--Fock theory and is responsible for the singlet-triplet energy gap in electronic excitations. In the case of methods based on the $\Delta\mathrm{SCF}$ approach, a spin-broken determinant is computed, whose energy is the arithmetic average between the singlet and the triplet states.
  
   \item Analyze the dependence of the CT descriptors on the distance separating the donor and acceptor moieties in space. In the limit of CT happening over a large distance, the magnitude of CT descriptors with distance units is expected to scale proportional to the distance between the centers of electronic charge of the monomers, while the adimensional descriptors should stay constant.
   This is a crucial point when assessing the performance of OO-DFT methods, as we analyze and explain later in the manuscript.
  
\end{itemize}

\section{Excited-state electronic-structure methods for the prediction of intermolecular charge-transfer}

Before briefly reviewing electronic-structure methods for the description of ICT, it is necessary to point out a set of requirements necessary to describe CT properly in a supramolecular system. We take Clever's cages\cite{han2014self} as a prototypical supramolecular system for which we aim to compute CT properties. In these supramolecules, the chromophore units are integrated in banana-shaped ligands forming the walls of the cage. By means of spectro-electrochemical methods\cite{frank2016light} and time-resolved ultrafast IR spectroscopy\cite{ahrens2017ultrafast}, it has been deduced that charge-transfer states are formed upon excitation, where one electron is transferred from the donor moiety (effectively oxidizing it) to an acceptor moiety, which is reduced. Photo-induced CT involving a chromophore integrated into the structure of a supramolecular cage as a donor and a molecular guest as an acceptor has been investigated previously\cite{ganta2022photoinduced} and has been shown to switch the charge-transfer properties of the material upon host-guest binding. On the one hand, the application of highly accurate, wavefunction-based electronic-structure methods to such a system is a rather prohibitive task; on the other hand, TD-DFT is capable of achieving accuracy in the range of tenths of eV for the prediction of excitation energies\cite{laurent2013td}. OO-DFT methods\cite{hait2020excited, hait2021orbital} display similar accuracy\cite{hait2020highly}, also for cases known to be problematic for general-purpose TD-DFT like core-excited states or doubly-excited configurations. A requirement for the predicted ICT excitation energy $E_{\mathrm{CT}}$ is to follow approximately the trend defined by Mulliken's formula\cite{mulliken1952molecular} for the distance dependence of the CT energy:

\begin{equation}\label{eq: mulliken}
    E_{\mathrm{CT}}=IP_{\mathrm{D}} - EA_{\mathrm{A}} - \frac{\mathrm{b}}{R_{\mathrm{DA}}},
\end{equation}

where $IP_{\mathrm{D}}$ is the ionization potential of the donor, $EA_{\mathrm{A}}$ the electron affinity of the acceptor, $R_{\mathrm{DA}}$ is the distance separating donor and acceptor, and $b$ is a system-dependent parameter with units of energy times distance. In the original formulation from Mulliken the parameter $\mathrm{b}$ is set to 1, whereas we introduced this parameter to fit the expression on the reference data later in the study. Since IP and EA are less sensitive to the interaction of the donor and acceptor moieties, the third term is the one that varies the most when the donor and acceptor are displaced in space. The CT energy is thus expected to increase with the donor-acceptor distance, approaching the $IP_{\mathrm{D}}-EA_{\mathrm{A}}$ asymptote. In large supramolecular systems, the distance between the donor and acceptor units can vary significantly, \textit{e.g.} bending vibrations involving the angle formed by a coordinated metal ion with the ligands constituting the walls of Clever's cages are a characteristic vibration of the supramolecular structure, important for the guest binding and release events\cite{juber2021thermodynamic}. Because of that, a fundamental requirement for an electronic-structure method to produce accurate ICT excitation energetics is that the correct asymptotic behavior is reproduced by the computed CTs for varying donor-acceptor distances.

Multiple studies\cite{baer2010tuned, prokopiou2022optimal, klawohn2020self} aimed to identify the adequate exchange-correlation functional form and parametrization for the accurate prediction of CT energies. In this work, we focus on range-separated hybrid exchange-correlation functionals (XCFs). We benchmark TD-DFT\cite{runge1984density} within the TDA\cite{hirata1999time} and compare the results to the maximum overlap method\cite{gilbert2008self} (MOM), initial MOM\cite{barca2018simple} (IMOM), and squared-gradient minimization\cite{hait2020excited} (SGM) DFT\cite{sham1966one} methods, which all belong to the class of OO-DFT excited-state electronic-structure methods (also known as $\Delta$SCF methods).
In addition to standard TDA calculations, we applied the non-empirical optimal tuning procedure proposed by Baers \textit{et al.}\cite{baer2010tuned} and implemented in Q-Chem\cite{epifanovsky2021software} for various XCFs to test their performance on ICT excitations. We also applied the Z-vector method\cite{furche2002adiabatic} to the TDA excited-state densities to obtain relaxed excited-state densities\cite{ronca2014density}, using the libwfa\cite{kimber2020toward} interface integrated in Q-Chem 6. \\

While TD methods analyze an approximation of the response function of the system to an external field to retrieve the excitation energies, the OO-DFT approach variationally optimizes the excited-state density, much like SCF algorithms do for the ground state. The key difference to ground-state electronic-structure theory is that the ground-state energy is a minimum of the electronic energy hypersurface in the occupied-virtual MO rotations, while excited open-shell determinants are saddle points. Convergence of saddle points on hyperdimensional surfaces instead of minima is a far more complicated task, so the SCF algorithms need to be adjusted to converge excited states. Two major approaches were proposed over the last 20 years, based on the (geometric) direct minimization ((G)DM)\cite{van2002geometric, schmerwitz2023calculations} SCF algorithm. 
Most OO-DFT methods exploit direct minimization (DM) methods, changing either the minimized objective function or the Fock matrix construction step in the SCF algorithm. 
In the Fock matrix construction step of a ground-state SCF algorithm, the new MOs are occupied according to the Aufbau principle which is not desirable for OO-DFT SCF algorithms, as it would mean re-coupling the excited electrons to the ground-state configuration. The MOM and IMOM methods therefore change the Fock matrix construction step to occupy the MOs that have the maximum overlap with a reference MO set, which can be either fixed (IMOM) or updated during the calculation (MOM). An alternative excited-state SCF algorithm is SGM, which works by minimizing the square of the gradient of the electronic energy in the MO rotations through a Newton-like optimizer. Since OO-DFT methods use local optimization methods, their convergence is highly sensitive to the initial guess. In our experience with the application of OO-DFT methods to the calculation of ICT excitations, using the restricted closed-shell ground state MOs is not always the optimal choice. An alternative guess for calculating low-lying CT excitations is to use the ground-state MOs from an unrestricted calculation for a mono-cationic system and add one electron to the target virtual orbital.
All OO-DFT methods investigated compute a spin-broken determinant, which is neither a singlet state nor a triplet state but whose energy lies halfway between the singlet and triplet energy. Usually, it is not recommended to use such an ansatz to calculate singlet excited state energies, as it is not a spin eigenfunction and typically yields a wrong energy. However, for CT excitations, the singlet-triplet energy gap is small and in the extreme case of pure ICT excitations, the singlet-triplet energy gap is negligible. Hence, it is not necessary to use a spin-purification formula to retrieve the singlet energy (see the SI for an extensive discussion).

Highly accurate reference excitation energies were computed using the EOM-CCSD(fT) method\cite{krylov2008equation}. They serve as reference data in this benchmark study alongside data points extracted from the literature computed with other coupled-cluster methods or second-order perturbation theory corrections, and experimental measurements (please see the SI for a detailed discussion of the reference data used).

\section{Calculation of the CT energy and CT descriptors for a small model system: ammonia-fluorine dimer}

According to Mulliken's formula (eq. \ref{eq: mulliken}), the ICT excitation energy is expected to change by displacing the donor (D) and acceptor (A) units in space. We require CT descriptors for all distances to distinguish ICT from intra-molecular CT and local excitations. In this section, we will test the abovementioned requirements for several electronic-structure models and compare to highly accurate benchmark data from wavefunction theory computed for a small model system, the ammonia-fluorine dimer, over a range of donor-acceptor separations (3.5, 4.25, 5, 8 and 10 \r{A}). The excited-state electronic-structure reference data is computed with the EOM-CCSD method\cite{krylov2008equation} and the cc-pVTZ basis set\cite{dunning1971gaussian}, yielding rather accurate electron densities and excitation energies. The excitation energies are then adjusted, computing the perturbative contribution from triple-excited configurations, using the EOM-CCSD(fT) method\cite{manohar2008noniterative} implemented in Q-Chem\cite{epifanovsky2021software}. We will first analyze the trends observed for the reference data and the performance of density-based CT descriptors on the accurate EOM-CCSD densities. Then, the performance of DFT-based excited-state methods using the LRC-$\omega$PBE\cite{rohrdanz2008simultaneous} XCF and the def2-TZVP\cite{weigend2005balanced} basis set is discussed.

\begin{figure}
  \centering
  \includegraphics[height=.33\linewidth]{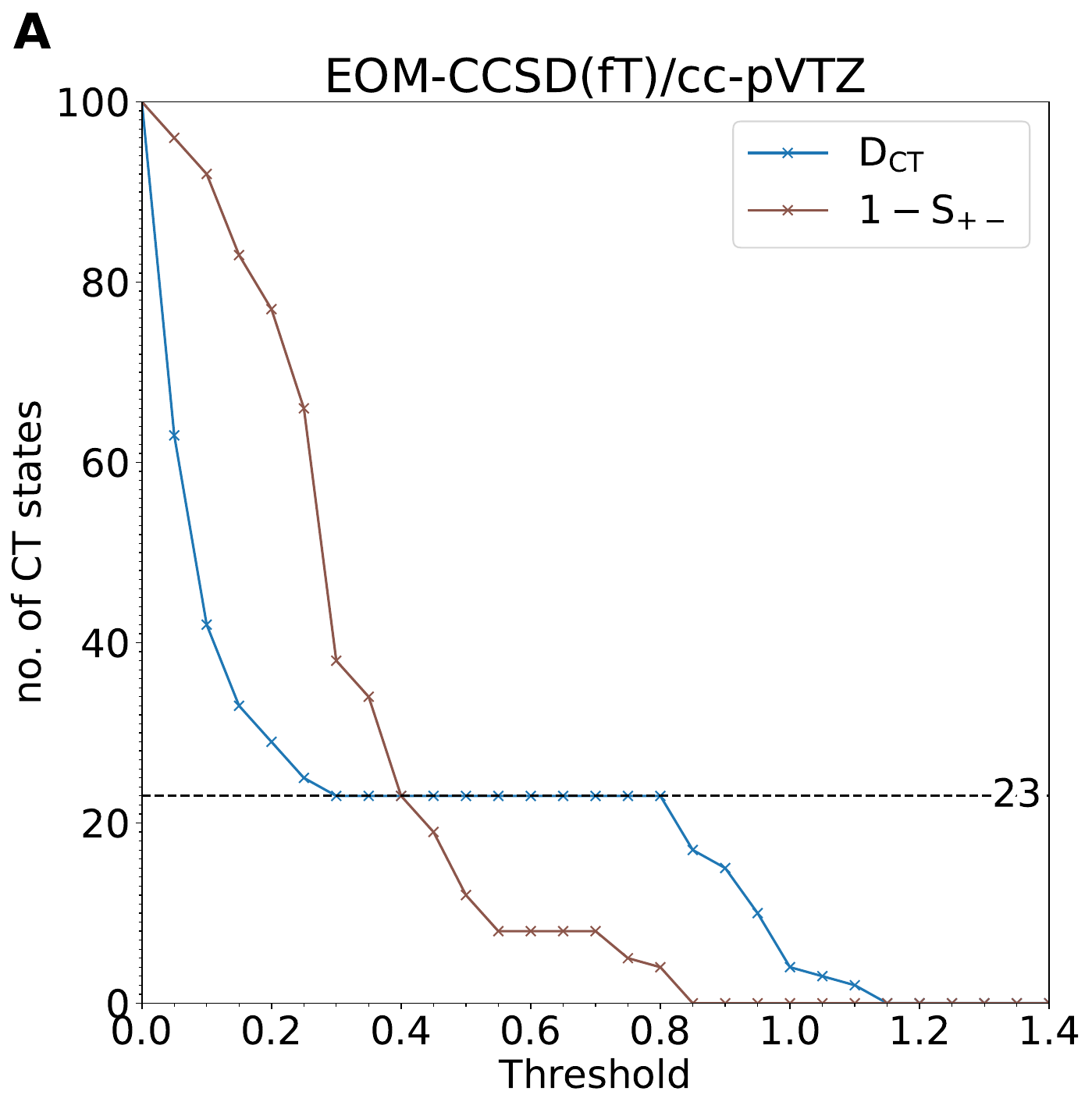}\hfill
  \includegraphics[height=.33\linewidth]{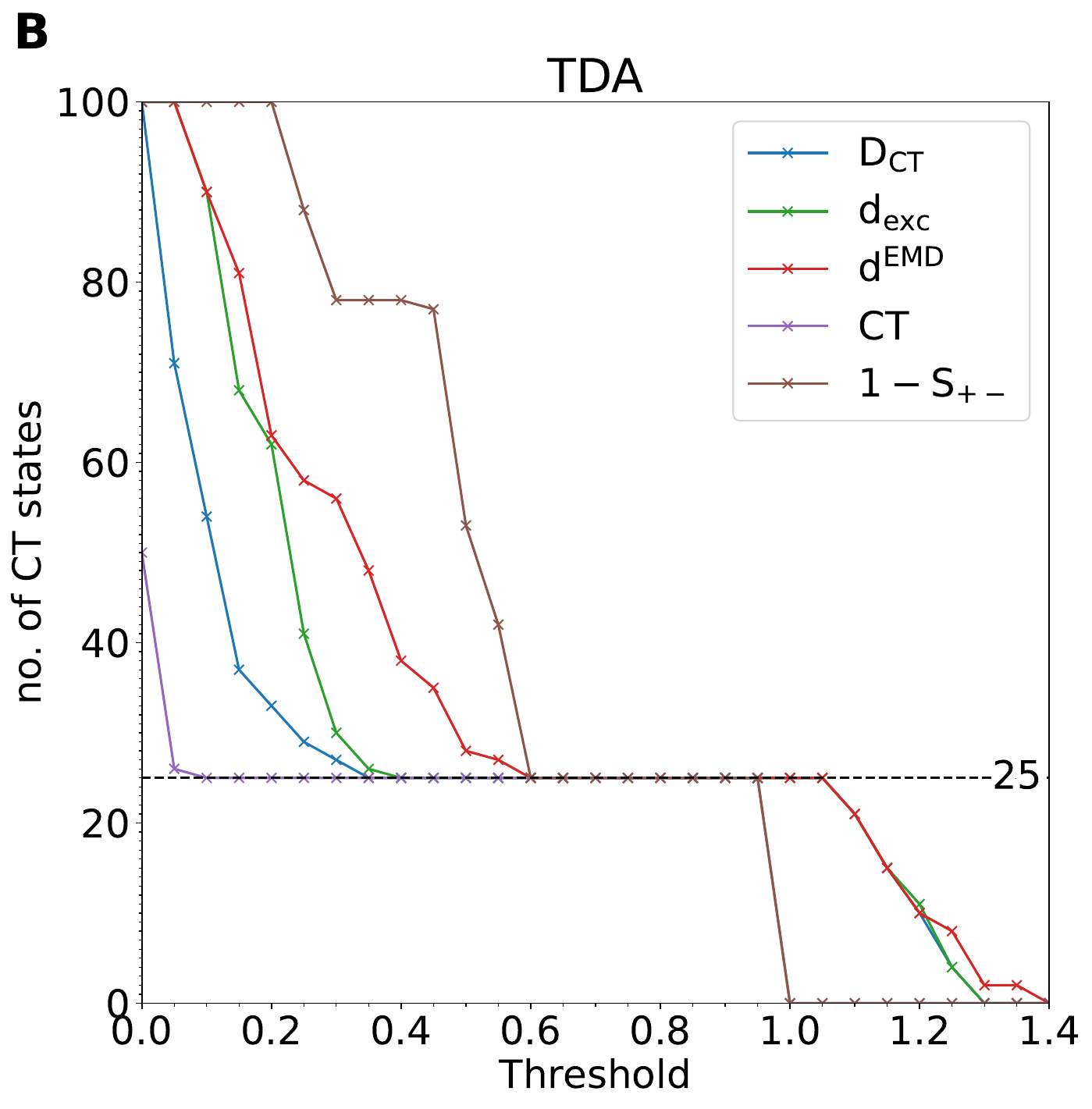} \hfill
  \includegraphics[height=.33\linewidth]{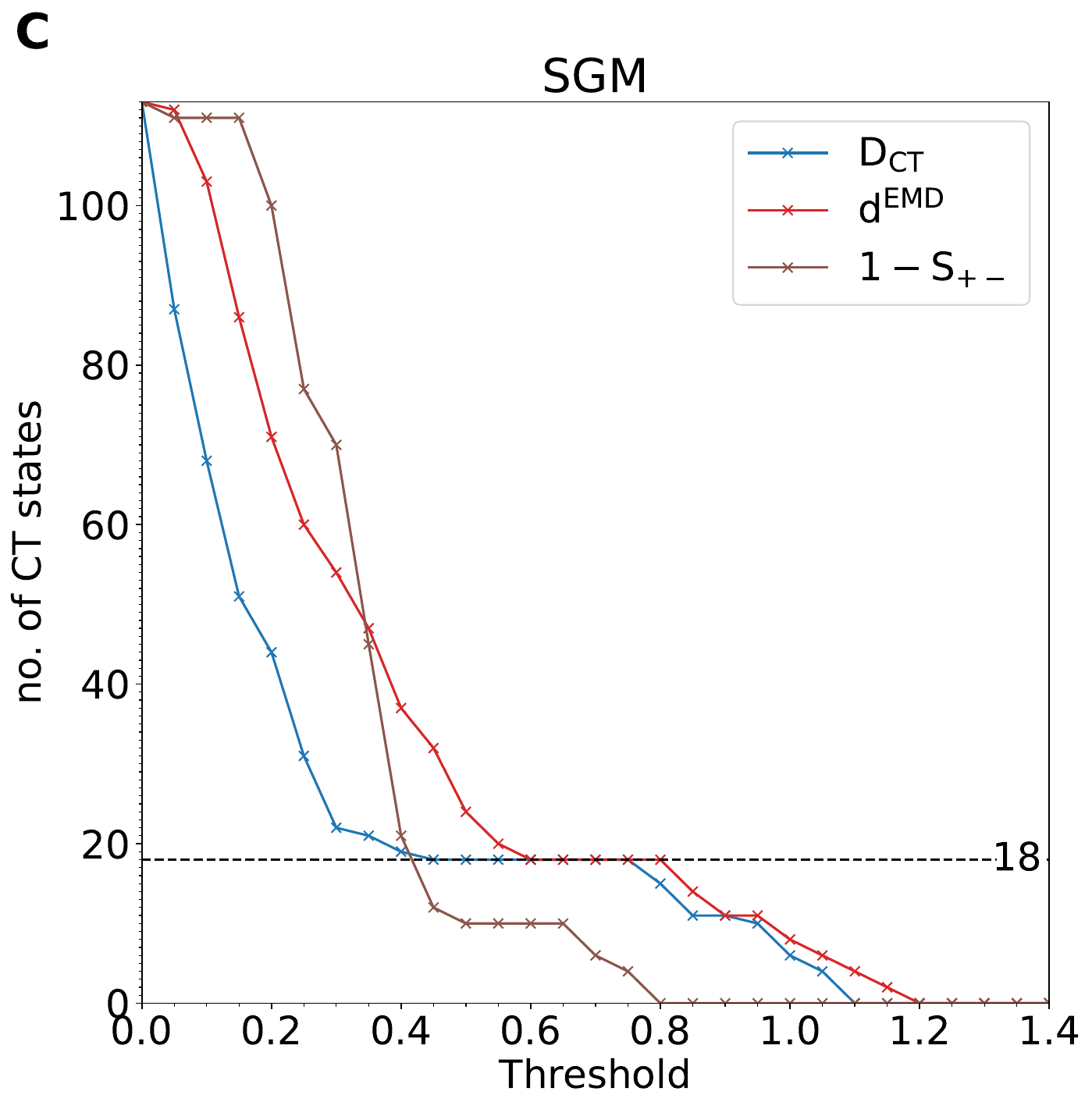}
  \caption{Number of identified CT states vs. assignment threshold for the EOM-CCSD(fT) (panel A), TDA (panel B), and SGM (panel C) methods, respectively. For a definition of the assignment threshold, see the main text.}
  \label{fig:spottedCT}
\end{figure}

%The capabilities of the various CT descriptors to identify ICT excitations are analyzed. 
The 20 lowest-lying excitations are computed with each method, leading to a total of 100 single-point excited states for the EOM-CCSD(fT) and TDA methods for the five donor-acceptor distances.
%The OO-DFT excited-state electronic-structure methods tested are SGM, IMOM, and MOM. 
The SGM, IMOM, and MOM calculations are set up using the closed-shell ground-state MOs as the initial guess, and all configurations with a weight greater than 0.3 in the CIS-like wavefunction from the TDA calculation are recomputed with the various OO-DFT calculations. All OO-DFT calculations employ unrestricted open-shell orbitals. Due to the selected threshold on the amplitudes, each CIS wavefunction can be dominated by up to three excited configurations, each potentially a CT excitation. This is the reason why in panel C of Figure \ref{fig:spottedCT} the total number of states exceeds 100, as more than 1 OO-DFT calculation was performed per each TDA state, on average. Additionally, it is essential to underline that the number of CT states identified by the descriptors can vary when comparing TDA to the other OO-DFT methods. 
On top of that, OO-DFT calculations are frequently prone to variational collapse. This means the number of CT states in the OO-DFT calculations varies for different methods or initial guesses. Hence, the number of assigned CT states can change substantially for the SGM, IMOM, and MOM results because some configurations with strong CT character in the TDA results converge to local excitations or vice versa. We analyze the ability of CT descriptors to identify convergence failure for CT excitations in OO-DFT methods.
Since TD-DFT produces a CIS-like wavefunction, both wavefunction-based and density-based CT descriptors can be calculated. In Figure \ref{fig:spottedCT}, we plot the number of identified CT excitations against the selected assignment threshold spanning a range from 0 to 1.4 for each method. For descriptors with distance units, the assignment threshold is defined as the ratio between the CT descriptor and the donor-acceptor distance $R_{\mathrm{DA}}$, while no such rescaling is necessary for dimensionless CT descriptors. Since $S_{+-}$ is expected to vanish with increasing $R_{\mathrm{DA}}$, CT states are detected when $1-S_{+-}$ is greater than the assignment threshold; all other parameters assign a CT state when they are greater than the threshold.
\begin{equation*}
  \text{State is } \begin{cases}
    \text{intermolecular CT}, & \text{if $\ \dfrac{D_\mathrm{CT}}{R_{\mathrm{DA}}},\ \dfrac{d_\mathrm{exc}}{R_{\mathrm{DA}}},\ \dfrac{d^\mathrm{EMD}}{R_{\mathrm{DA}}},\ (1-S_{+-}) >\ \mathrm{thresh.}$}\\
    \text{local excitation}, & \text{otherwise}.
  \end{cases} \\
\end{equation*}

\vspace{0.5cm}

All wavefunction-based parameters are computed with TheoDORE, and the molecular structure is automatically divided into two fragments with the OpenBabel\cite{o2011open} package.
For all DFT methods, density-based parameters are computed. $D_{\mathrm{CT}}$ and $S_{+-}$ are computed with Multiwfn\cite{lu2012multiwfn} by processing density cube-files generated with the electronic-structure program. To compute the EMD-based parameters, a development branch of Q-Chem was used to write the density data to an SG-1 quadrature grid\cite{gill1993standard}. The resulting density was fitted to a ``(19,26)” grid with the ChargeEMD\cite{wang2023earth} package, where 19 is the number of grid points in the radial part, and 26 is the number of grid points in the angular part for each atom. The ChargeEMD package then uses this coarser grid to solve the EMD problem.

As reported in panel A of Figure \ref{fig:spottedCT}, the number of ICT states identified by the $D_\mathrm{CT}$ diagnostic for the EOM-CCSD(fT) calculations is constant over a large range of specified assignment thresholds, leading to a plateau with 23 CT states. This is not the case for the $S_{+-}$ descriptor, where no such plateau is observed signaling an undesired strong dependence of the CT assignment on the threshold. All descriptors consistently detect 25 CT excitations for the TDA calculations (panel B). The descriptor giving the most consistent results over the largest assignment threshold range is $CT$, which is either close to 1 or 0, depending on whether or not the corresponding state is a CT excitation.
$D_\mathrm{CT}$ and $d^\mathrm{EMD}$ consistently match 21, 22, and 18 CT excitations for MOM, IMOM, (see SI for both plots), and SGM (panel C), respectively. 
For all OO-DFT methods, $S_{+-}$ shows the same undesired behavior as for the EOM-CCSD densities, limiting its applicability. Since the density-based $D_\mathrm{CT}$ descriptor shows good agreement with other CT descriptors and minor sensitivity to the assignment threshold, it will be used exclusively in the remainder of our study.

In Figure \ref{fig:lowestCT}, we compare only the lowest-energy CT excited state with the reference data. As predicted by Mulliken's formula (eq. \ref{eq: mulliken}), the EOM-CCSD(fT) CT excitation energy (panel A) increases monotonously upon displacing the donor and acceptor units. While the standard TDA results (blue line) capture the asymptotic trend of the CT excitation energy, it is strongly red-shifted. In this example, the error is greater than 3~eV, making the predicted energy of little use for the rational design of supramolecular assemblies. However, TDA benefits from the optimal tuning procedure of the XCF proposed by Baer \textit{et al.}\cite{baer2010tuned} (denoted with TDA-OT in Fig \ref{fig:lowestCT}, red line), as the error on the excitation energies reduces by about 50~\% to 1.5 eV while retaining the correct asymptotic trend. For the OO-DFT methods, SGM shows consistent asymptotic behavior, and the predicted CT excitation energy is red-shifted by only about 0.5 eV, except for the data point calculated at 5\r{A} separation.

The maximum-overlap methods (IMOM and MOM) show errors similar to those of SGM. For the OO-DFT methods there are obvious convergence problems at short donor-acceptor distances, \textit{e.g.} the data point produced with MOM at 3.5 \r{A} is the result of convergence to a state with smaller CT character (see panel B), and IMOM computes the same excitation energy. The SGM algorithm is not immune to similar errors (see abovementioned data point at 5\r{A} separation) due to convergence to other roots or to regions of the MO rotations hypersurface that are not saddle points of the electronic energy, so they do not belong to the electronic structure of excited configurations\cite{burton2022energy}. The data points computed using OO-DFT methods and the default $\omega$ parameter for large separation distances are indistinguishable from one to another, blue-shifted by only about 0.5 eV compared to the reference.

\noindent\begin{minipage}{\linewidth}
\centering
 \includegraphics[width=\textwidth]{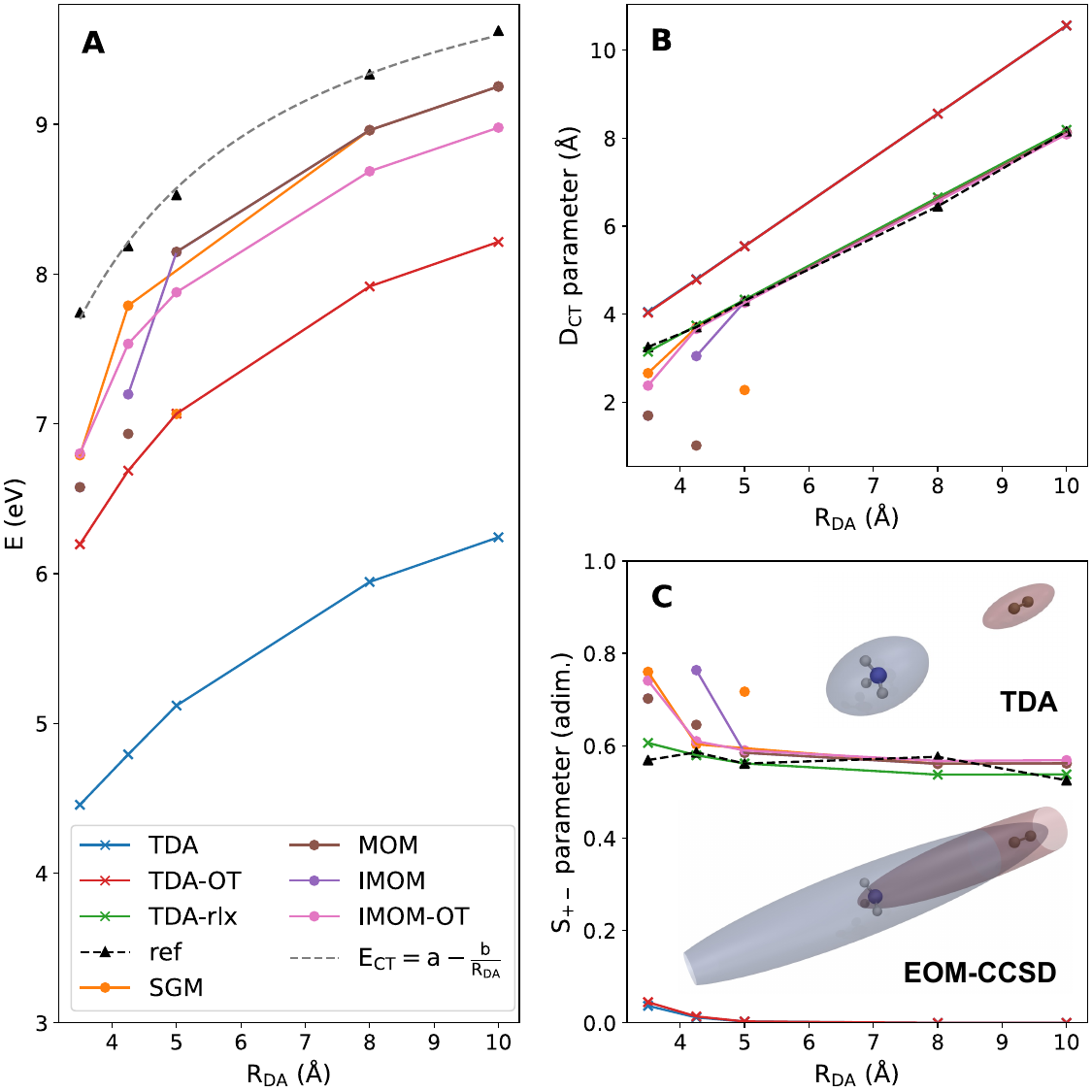}
 \captionof{figure}{Panel A: CT excitation energy belonging to the lowest-lying CT state of the ammonia-fluorine dimer vs. donor-acceptor separation distance (\r{A}) computed with the EOM-CCSD(fT) (ref.), TDA, optimally-tuned TDA (TDA-OT), IMOM, MOM, and SGM methods, respectively. Data points characterised by $D_\mathrm{CT} > \frac{1}{2} R_\mathrm{DA}$ are connected with a solid lines. The reference data is fitted to Mulliken's equation for the CT excitation energy (eq. \ref{eq: mulliken}), using the 2-parameter expression $E_\mathrm{CT} = a - \frac{b}{R_\mathrm{DA}}$, where $a = IP_{\mathrm{D}} - EA_\mathrm{A}$ and $b$ is a system-specific parameter with unit of energy times distance. Panel B: $D_\mathrm{CT}$ parameter (\r{A}) computed on the lowest-lying CT state vs. donor-acceptor separation distance (\r{A}) computed with the EOM-CCSD(fT) (ref.), TDA, optimally-tuned TDA (TDA-OT), relaxed TDA (TDA-rlx), IMOM, MOM, and SGM densities. Panel C: $S_{+-}$ diagnostic computed for the lowest-lying CT state vs. donor-acceptor separation distance (\r{A}). Inset: isosurface plot of the normalized $C_+$ and $C_-$ ellipsoids produced by Multiwfn, using TDA and EOM-CCSD densities at 10 \r{A} separation. The isovalue is $\pm$0.0001 for $C_+$ and $C_-$.}
 \label{fig:lowestCT}
\end{minipage} 

 While the TDA method benefits greatly from tuning the range-separation parameter of the XCF, IMOM gives a larger error of around 1 eV when using the optimal $\omega$ parameter, (pink line in Figure \ref{fig:lowestCT}). Due to the more stable convergence behaviour over SGM, we will use IMOM for all OO-DFT calculations in the following part of the study.

The $D_{\mathrm{CT}}$ diagnostic (panel B) increases linearly with the donor-acceptor separation $R_{\mathrm{DA}}$ for the reference results. The OO-DFT calculations that converged reliably onto true ICT states (circle markers connected by the solid lines), as well as the relaxed densities calculated with the Z-vector method (green line, cross markers), are characterized by a $D_\mathrm{CT}$ diagnostic very close to the reference. In stark contrast, the unrelaxed TDA densities (red and blue plots, cross marker) return a more pronounced CT character than the reference EOM-CCSD density.
As expected, $S_{+-}$ (panel C) gives 0 overlap for the density depletion and accumulation volumes in the CT state for the unrelaxed TDA densities, while its value stays rather large for all methods providing an accurate density. In general, density-based CT descriptors are enhanced when applying the TDA: the molecular orbitals are unchanged in the excited state, so the charge separation does not polarize the approximate excited-state density.
On the other hand, for the reference calculations as well as for orbital-optimized methods, the electrons belonging to the acceptor unit are attracted toward the cationic donor unit, and the resulting density is accumulated at shorter distances. Accordingly, the $C_+$ and $C_-$ ellipsoids are elongated along the direction of the Coulomb attraction, to the point where they exceed the volume of the cube file used for the electron density calculation. This is observed in the inset in panel C, where the localized ellipsoids from the TD-DFT density are displayed alongside the elongated, cut-off ellipsoids computed using the accurate density.

In conclusion, the density-based CT descriptors calculated with the OO-DFT densities agree substantially better with the reference than the TDA results for the lowest-lying ICT excitation in the ammonia-fluorine dimer. Relating density-based CT descriptors like $D_{\mathrm{CT}}$ and $d^{\mathrm{EMD}}$ to $R_{\mathrm{DA}}$ helps to identify convergence failure of the OO-DFT calculations.

\section{Benchmarking the distance dependence of ICT excitations}

We benchmarked the performance of DFT-based excited-state methods to accurately describe ICT on a variety of small molecular dimers. To generate accurate reference data, we computed the 20 lowest-lying singlet excited states of the beryllium-fluorine dimer and the tetrafluoroethylene-ethylene dimer with EOM-CCSD(fT) and the cc-pVTZ basis at 3.5, 4.25, 5, 8 and 10 \r{A} distance between the molecular center of mass. All states with a $D_\mathrm{CT}$ greater than half of the donor-acceptor distance $R_\mathrm{DA}$ were classified as ICT states. The same number of lowest-lying excitations was computed with the TDA method and IMOM using the LRC-$\mathrm{\omega}$PBE XCF and the def2-TZVP basis. In addition, a set of data points with the lowest-lying ICT excitations of the ammonia-nitrous acid dimer computed with the SA2-MS-CAS(2,2)PT2 method and the jul-cc-pVTZ basis at nine separation distances was taken from the literature\cite{ghosh2015multiconfiguration}. This system was included in our reference dataset, which we will refer to as the $R_\mathrm{DA}$-dataset from here on.

\begin{table}[h!]
\centering
\begin{tabular}{||c || c c || c c || c c||} 
\hline
\multicolumn{1}{||c||}{} &\multicolumn{2}{|c||}{TDA} &\multicolumn{2}{|c||}{TDA-OT} &\multicolumn{2}{|c||}{IMOM} \\
 \hline
 Dimer & MSE & MSV & MSE & MSV & MSE & MSV \\
 \hline\hline
 $\mathrm{Be-F_2}$ & -2.71 & 1.21 & -1.71 & 0.57 & -1.23 & 0.01 \\ [0.5ex] 
 $\mathrm{NH_3-F_2}$ & -3.36 & 0.4 & -1.45 & 0.32 & -0.53 & 0.08 \\ [0.5ex] 
 $\mathrm{NH_3-HNO_3}$ & -1.64 & 0.01 & -0.22 & 0.00 & 0.07 & 0.00 \\ [0.5ex] 
 $\mathrm{C_2F_4-C_2H_4}$ & -0.74 & 0.08 & -0.74 & 0.08  & -0.16 & 0.12 \\ [0.5ex] 
 All & -2.09 & 1.52 & -1.03 & 0.47 & -0.55 & 0.49 \\ [0.5ex] 
 $D_\mathrm{CT} > 0.5\ R_\mathrm{DA}$ & -2.13 & 1.55 & -0.93 & 0.42 & -0.27 & 0.14 \\ [0.5ex] 
 \hline
\end{tabular}
\caption{Mean signed error (MSE) and mean signed variance (MSV) on the ICT excitation energy for each dimer system scan investigated, using the various DFT-based excited-state electronic structure methods, the LRC-$\mathrm{\omega}$PBE XCF and the def2-TZVP basis. Entries in the column \textit{All} correspond to a linear fit of all datapoints combined. All values reported in eV unit.}
\label{table: dimers}
\end{table}

Table \ref{table: dimers} summarizes the performance of DFT-based excited-state methods for the prediction of ICT excitation energies over the $R_\mathrm{DA}$-dataset. For the tetrafluoroethylene-ethylene dimer, many excited states show intermediate $D_\mathrm{CT}$ values, close to half $R_\mathrm{DA}$, especially at short donor-acceptor separation (3.5, 4.25, 5 \r{A}). For all of these states, the examination of the CIS-like wavefunction produced by TDA showed no single dominant excited configuration; instead, several singly-excited determinants exhibited similar amplitudes. The data points belonging to the three lowest-lying ICT states were selected using information from adjacent separation distances. Amongst the states with significant $D_{\mathrm{CT}}$ character, the data points reproducing the monotonous trend predicted by Mulliken's formula (eq. \ref{eq: mulliken}) were classified as ICT, such that the energy of each CT root was increasing by displacing donor and acceptor far in space.

TDA (column 2) tends to underestimate the energy of ICT states by several eV. The generally poor performance on the whole dataset (5th row) is manifested in a large variance, meaning the error is not a simple offset in energy, but is due to systematic scattering in the range of several eV in the worst cases, like the beryllium-fluorine (first line of Table \ref{table: dimers}). In the case of organic systems like the tetrafluoroethylene-ethylene dimer (second line), the MSE drops below 1 eV. This is reasonable since the LRC-$\omega$PBE XCF was parametrized with the Minnesota Database/3\cite{lynch2003robust}, which is mainly composed of small organic molecules, and a set of ionization potentials, electron affinities, and excitation energies\cite{rohrdanz2008simultaneous}. In fact, LRC-$\omega$PBE has been shown to successfully estimate the excitation energy of CT states in polycyclic aromatic hydrocarbons\cite{richard2011time}. Still, the variance increases dramatically when the data points from all molecular dimers are combined. This means the method's performance is system-dependent and cannot be corrected by a universal shift or offset. In  supramolecules, multiple ICT excitations are possible, and electronic structure methods must be capable of computing all ICT excitations with negligible or small transferable error. TDA with the default LRC-$\omega$PBE XCF is clearly not capable of that, and we therefore investigated the performance of other methods.

Column 3 in Table \ref{table: dimers} shows the performance of the TDA method with an optimally-tuned range-separation parameter $\omega$ in the LRC-$\omega$PBE XCF. It has been shown that the range-separated BNL XCF\cite{livshits2007well} proved to benefit from optimal tuning for the description of ICT excitations in small organic molecules\cite{stein2009reliable} and optimal tuning also significantly improves the performance of LRC-$\omega$PBE with the TDA method in our study, lowering the MSE by up to 1.5 eV. The fact that the tetrafluoroethylene-ethylene dimer is largely unaffected can be explained by considering that the default $\omega$ parameter was selected by benchmarking against data including redox properties of organic systems. Yet, the error and the variance of the combined dataset remain significant, and hence the results from TDA cannot be directly compared to experimental data for ICT.

In the last column of Table \ref{table: dimers}, we show the results obtained with IMOM for the same $R_\mathrm{DA}$-dataset. The default value of the $\omega$ parameter was used in these calculations since, as we have observed for the ammonia-fluorine dimer in the previous section, optimal tuning does not yield an improvement for ICT states with IMOM. Some calculations on the tetrafluoroethylene-ethylene dimer converged on electron densities with small $D_\mathrm{CT}$ and electronic energies far from the reference, hinting at a numerical issue for this high-dimensional problem\cite{selenius2023orbital}, and we excluded them from this benchmark study. Without these data points, IMOM produces the best results for the calculation of ICT excitation energies for all molecular dimer systems, with MSE in the order of tenths of eV, with the exception of the beryllium-fluorine system. This system proved to be the most challenging to compute using an RSH XCF and is not representative of the ICT states that are likely to be formed in supramolecular assemblies. For this reason, the error statistics excluding this system are shown in the last line of Table \ref{table: dimers}, which, in our opinion, gives the most representative picture for the error of all DFT-based methods discussed here.

In Figure \ref{fig: RDA stats}, we show the explicit data points and linear correlations for all methods. Clearly, IMOM (panel C) produces the smallest errors with little variance. It is also evident that optimally-tuned LRC-$\omega$PBE improves considerably over the results obtained with the default $\omega$ parameter.

\begin{center}
    \includegraphics[width=\textwidth]{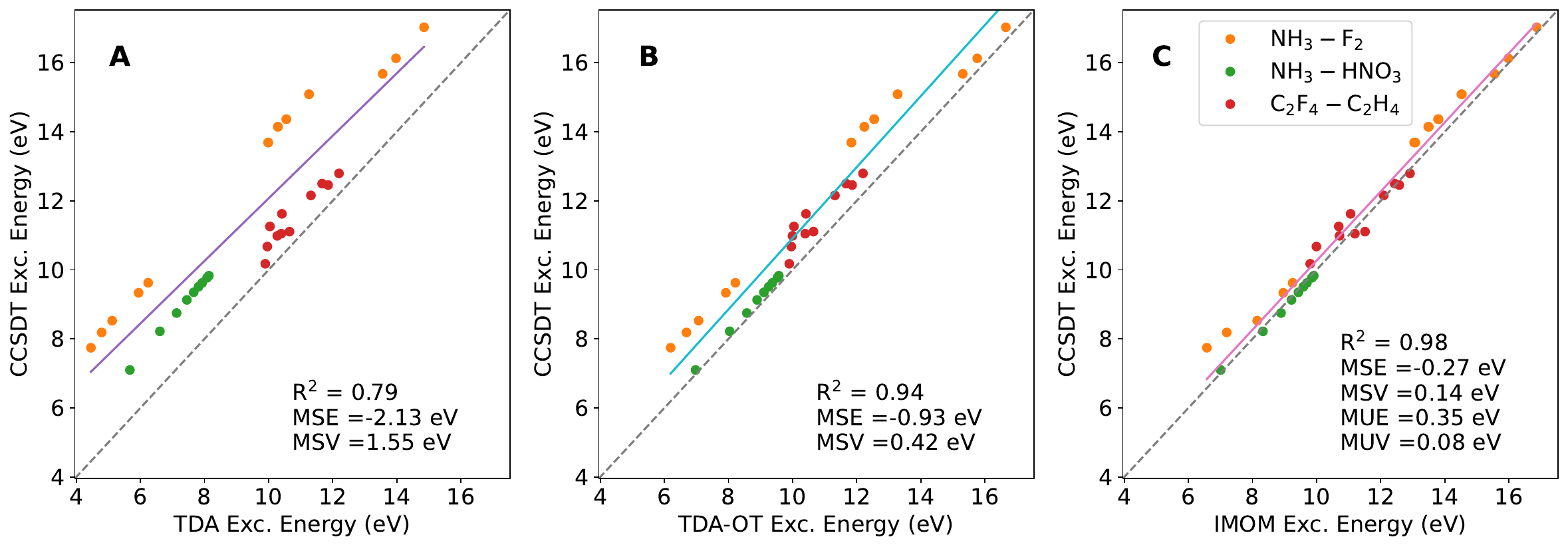}
 \captionof{figure}{Excitation energy to the low-lying ICT states in the $R_{\mathrm{DA}}$-dataset. The data points belonging to the Beryllium-fluorine dimer were excluded from this plot. Dashed grey line: target trend, where the reference and the DFT energy are the same. Solid lines: linear fit. The color-coding highlights data points belonging to the molecular dimers in the dataset.}
 \label{fig: RDA stats}
\end{center}

\section{Benchmarking TDA and OO-DFT methods in a larger set of small molecular dimers}

While the $R_\mathrm{DA}$-dataset includes CT excitation energies for several donor-acceptor separation distances, the dimers included in this dataset are chemically not very diverse. We therefore conducted an extensive literature search to test the selection of TDA and OO-DFT methods for a diverse set of molecular dimers with ICT states\cite{zuluaga2020benchmarking, bokhan2015explicitly, masnovi1984electron, hanazaki1972vapor, yu2011accurate, musial2011charge, zhao2006density, kozma2020new, ghosh2015multiconfiguration, solovyeva2014describing, dutta2018exploring, tawada2004long, plasser2014new, stein2009reliable, manna2018prediction}. We selected the most accurate electronic-structure results obtained with the largest basis sets in our reference data. More information on this benchmark set and the selected reference methods can be found in the SI. We followed the same computational protocol and ICT classification scheme as for $R_\mathrm{DA}$-dataset.
%Following the protocol used for the $R_\mathrm{DA}$-dataset, TDA calculations were performed for the 20 lowest-lying singlet excited states with the LRC-$\omega$PBE XCF and def2-TZVP basis. 
%The donor-acceptor separation distance $R_\mathrm{DA}$ was measured as the distance between the centers of mass of the donor and acceptor molecules, and states characterized by a $D_\mathrm{CT}$ value greater than $\frac{1}{2}R_\mathrm{DA}$ were classified as ICT excitations. Then, IMOM calculations were performed for each excited configuration with an amplitude greater than 0.3 in the CIS-like wavefunction of all ICT states computed with the TDA method. 
In Figure \ref{fig: large dataset}, the correlation of the DFT results with the reference data is plotted with specified color coding for different reference methods and data sources.

\begin{center}
    \includegraphics[width=\textwidth]{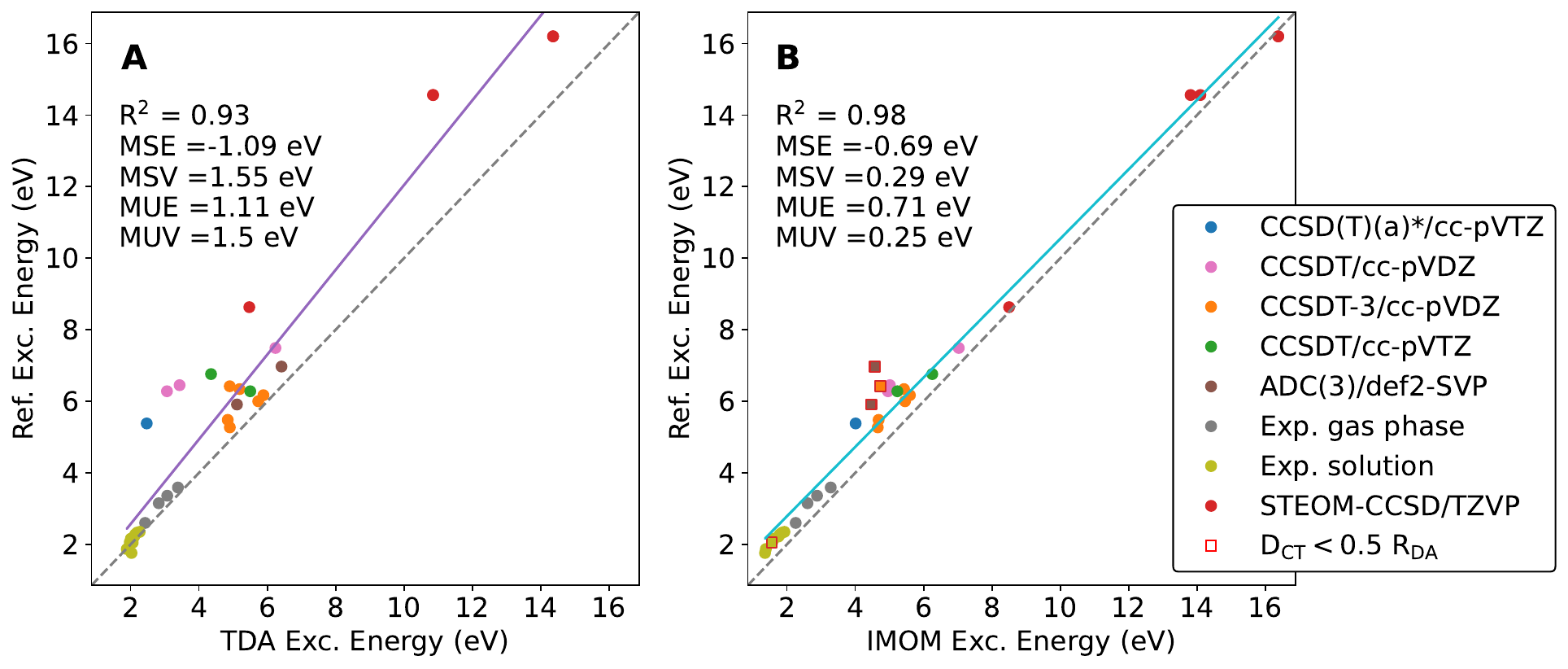}
 \captionof{figure}{Correlation plot of excitation energies of calculated low-lying ICT states with the literature data. Panel A summarizes the TDA results, whereas the IMOM results are summarized in Panel B. The dashed grey line corresponds to perfect agreement between our calculations and the reference date whereas the solid line is the result of a linear fit. Different reference results are color-coded according to the legend and outliers due to convergence to local excitations (identified by $D_\mathrm{CT} < \frac{1}{2} R_\mathrm{DA}$) are marked with a square in panel B.}
 \label{fig: large dataset}
\end{center}

Again, IMOM outperforms TDA with the default $\omega$ parameter, achieving both a significantly lower error and reduced variance. On top of that, IMOM yields much better results for large excitation energies. Both methods perform well on the experimental data for low-lying ICT states as well.\\ 

The goal of this study is not a thorough theory to experiment benchmark, which would come with its own challenges especially for dimers in solution, where solvent rearrangement has a major effect on the CT states. We rather seek to identify a suitable electronic-structure method that describes these states reliably at low computational cost. In that sense, a comparison to accurate theoretical reference data is most meaningful. In Figure \ref{fig: small dataset}, we hence compare the accuracy of TDA, OT-TDA and IMOM on a subset of our dataset, where we excluded experimental data and those data points where IMOM did not converge to a CT state ($D_\mathrm{CT} < \frac{1}{2} R_\mathrm{DA}$).

\begin{center}
    \includegraphics[width=\textwidth]{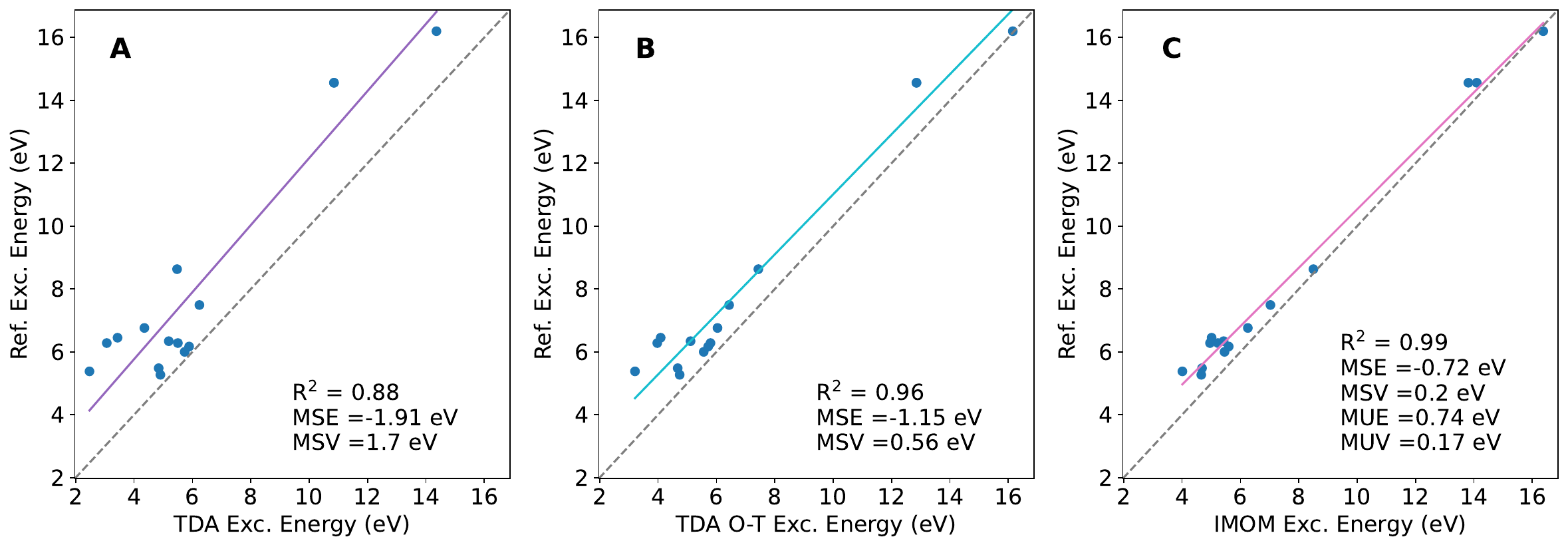}
 \captionof{figure}{Excitation energies of the low-lying ICT states obtained from the literature, including only accurate \textit{ab initio} reference data. The dashed grey line corresponds to perfect agreement between our calculations and the reference date whereas the solid line is the result of a linear fit. TDA results are displayed in panel A, OT-TDA results are displayed in panel B and panel C shows the results of the IMOM calculations.}
 \label{fig: small dataset}
\end{center}

As we have seen before, optimal tuning (panel B) considerably reduces the error compared to TDA, but IMOM shows the best overall performance. While the MSE on this subset of the dataset from the literature is still considerably high (an underestimation of the ICT excitation energy by 0.7 eV), the variance of the IMOM data is less than half that of the optimally-tuned TDA method. This points to a systematic and system-independent underestimation of the ICT energies. While the improvement obtained using the optimal-tuning procedure over the default $\omega$ parameter for the LRC-$\omega$PBE XCF is noteworthy, the variance is still quite large and the error is hence more system-dependent.

Although we demonstrated the limitations of TDA regarding the calculation of ICT excitations, the method plays a crucial role in the computational protocol employed to set up IMOM calculations for low-lying ICT states. In addition, the performance of TDA critically depends on the choice of an adequate XCF and basis set. Obviously, a plethora of XCFs were developed over the years using various combinations of functional form and parameterization. Therefore, we analyzed several XCF and basis set combinations on the reduced data set used in Figure \ref{fig: small dataset}, applying the optimal tuning procedure. The results are summarized in Tables \ref{table: bases} and \ref{table: xcfs}.

\begin{table}[h!]
\centering
\begin{tabular}{||c c c c c c c||} 
 \hline
 Method & XCF & Basis & MSE & MSV & MUE & MUV \\
 \hline\hline
 TDA & LRC-$\mathrm{\omega}$PBE & def2-SVP & -1.66 & 1.50 & 1.66 & 1.50 \\ [0.5ex] 
 TDA & OT LRC-$\mathrm{\omega}$PBE & def2-SVP & -0.53 & 0.50 & 0.77 & 0.18 \\ [0.5ex] 
 TDA & LRC-$\mathrm{\omega}$PBE & def2-TZVP & -1.75 & 1.57 & 1.75 & 1.57 \\ [0.5ex] 
 TDA & OT LRC-$\mathrm{\omega}$PBE & def2-TZVP & -1.14 & 0.46 & 1.14 & 0.46 \\ [0.5ex] 
 TDA & LRC-$\mathrm{\omega}$PBEh & def2-SVP & -1.92 & 1.39 & 1.92 & 1.39 \\ [0.5ex] 
 TDA & OT LRC-$\mathrm{\omega}$PBEh & def2-SVP & -0.64 & 0.67 & 0.91 & 0.23 \\ [0.5ex] 
 TDA & LRC-$\mathrm{\omega}$PBEh & def2-TZVP & -2.04 & 1.37 & 2.04 & 1.37 \\ [0.5ex] 
 TDA & OT LRC-$\mathrm{\omega}$PBEh & def2-TZVP & -1.07 & 0.61 & 1.09 & 0.57 \\ [0.5ex] 
 \hline
 IMOM & LRC-$\mathrm{\omega}$PBE & def2-TZVP & -0.72 & 0.20 & 0.74 & 0.17 \\ [0.5ex] 
 \hline
\end{tabular}
\caption{Mean signed error (MSE), mean signed variance (MSV), mean unsigned error (MUE), and mean unsigned variance (MUV) of ICT excitation energies computed using various methods, XCF, and basis set combinations. Error statistics are calculated with respect to the highly accurate quantum-chemical reference calculations subset also used in Figure~\ref{fig: small dataset}. All values are reported in eV.}
\label{table: bases}
\end{table}

Table \ref{table: bases} shows the dependence of the results on the basis set size for the TDA method, using both the default range-separation parameter and an optimally tuned value for the LRC-$\omega$PBE and LRC-$\omega$PBEh XCFs. Notably, the benefit from the optimal tuning procedure is more pronounced for the smaller def2-SVP basis set than for def2-TZVP, lowering the error by up to 1.2 eV. The MSE is similar to the one obtained with IMOM and a large basis set. However, the variance is rather large, indicating a more system-dependent performance. The inverted  basis set  dependence is, in principle, undesired, since it compromises the systematic improvement of results but has practical relevance since it motivates the use of small basis sets for cost-efficient calculations of ICT excitation energies.

\begin{table}[h!]
\centering
\begin{tabular}{||c c c c c||} 
 \hline
 XCF & MSE & MSV & MUE & MUV \\
 \hline\hline
 LRC-$\mathrm{\omega}$PBE\cite{rohrdanz2008simultaneous} & -1.64 & 1.42 & 1.64 & 1.42 \\ [0.5ex] 
 OT LRC-$\mathrm{\omega}$PBE & -0.45 & 0.62 & 0.78 & 0.18 \\ [0.5ex] 
 LRC-$\mathrm{\omega}$PBEh\cite{rohrdanz2009long} & -1.92 & 1.39 & 1.92 & 1.39 \\ [0.5ex] 
 OT LRC-$\mathrm{\omega}$PBEh & -0.64 & 0.67 & 0.91 & 0.23 \\ [0.5ex] 
 LRC-BOP\cite{song2007long} & -0.16 & 0.66 & 0.68 & 0.20 \\ [0.5ex] 
 OT LRC-BOP & -0.41 & 0.56 & 0.72 & 0.19 \\ [0.5ex]
 LC-VV10\cite{vydrov2010nonlocal} & -0.21 & 0.67 & 0.69 & 0.21 \\ [0.5ex] 
 OT LC-VV10 & -0.44 & 0.62 & 0.78 & 0.18 \\ [0.5ex] 
 LC-rVV10\cite{mardirossian2017use} & -0.21 & 0.67 & 0.69 & 0.21 \\ [0.5ex] 
 OT LC-rVV10 & -0.44 & 0.62 & 0.78 & 0.18 \\ [0.5ex] 
 CAM-B3LYP\cite{yanai2004new} & -1.98 & 1.31 & 1.98 & 1.31 \\ [0.5ex] 
 OT CAM-B3LYP & -0.87 & 0.70 & 0.89 & 0.67 \\ [0.5ex]
 $\mathrm{\omega}$B97M-V\cite{mardirossian2015mapping} & -1.07 & 0.71 & 1.07 & 0.71 \\ [0.5ex] 
 OT $\mathrm{\omega}$B97M-V & -0.31 & 0.40 & 0.58 & 0.15 \\ [0.5ex] 
 \hline
\end{tabular}
\caption{Mean signed error, mean signed variance, mean unsigned error, and mean unsigned variance of ICT excitation energies for several XCFs calculated with TDA and the def2-SVP basis with and without optimal tuning of the range-saparation parameter. Error statistics are calculated with respect to the highly accurate quantum-chemical reference calculations subset also used in Figure~\ref{fig: small dataset}. All values are reported in eV.}
\label{table: xcfs}
\end{table}

In Table \ref{table: xcfs}, we report error statistics for several common XCFs for the same reduced dataset employing TDA calculations and the small def2-SVP basis set. The MSE depends strongly on the XCF, but all optimally-tuned XCFs achieve sub-eV accuracy. In contrast, when the default $\omega$ is used, most of the XCFs perform poorly, except for LRC-BOP and some of the VV10-based XCFs. Neither LRC-BOP nor the VV10 family of XCFs were parametrized against CT energies, ionization potentials, or electron affinities. The VV10-based XCFs were previously shown\cite{vydrov2010nonlocal} to yield good performance on ground-state energies of molecular dimers, as they accurately compute van-der-Waals interactions but were not designed for the calculation of the redox properties. Isolated optimal tuning of the range-separation parameter is generally not advised for XCFs where all parameters were simultaneously optimized, like CAM-B3LYP and $\omega$B97M-V. Yet, the predicted ICT excitation energies benefit from tuning. This is to be expected, since improving the approximate IP and EA values produced by an XCF should also improve the ICT excitation energies, which follows directly from Mulliken's formula (eq. \ref{eq: mulliken}). The improvement is remarkable and we highlight the excellent performance of the optimally-tuned $\omega$B97M-V functional both in terms of error and variance for the small basis set.

\section{Conclusions}

We analyzed a diverse set of charge-transfer descriptors for ICT excitations in small molecular dimers, and found that especially $D_\mathrm{CT}$ is well-suited to reliably identify the CT character for these systems. We then compared several DFT-based excited-state methods with highly accurate reference data from wavefunction theory for ICT excitation energies. Among these, IMOM provides the most robust performance among the OO-DFT methods. A new dataset containing reference data with several donor-acceptor separations, termed $R_\mathrm{DA}$-dataset, was then established using the EOM-CCSD(fT) method and a cc-pVTZ basis. All DFT methods were then tested against the reference data in the $R_\mathrm{DA}$-dataset (see Table \ref{table: dimers}), using the LRC-$\omega$PBE XCF and the def2-TZVP basis set. Standard TDA calculations showed the worst performance, with both MSE and MSV significantly larger than 1 eV. The non-empirical optimal-tuning procedure from Baer \textit{et al.}\cite{baer2010tuned} was then applied, yielding a significant improvement over the results obtained with the default range-separation parameter. Yet, the performance of optimally-tuned TDA is not satisfactory for the general calculation of ICT states, as its variance is in the range of 0.5 eV. In stark contrast, IMOM showed satisfying performance, with sub-eV errors and a variance also on the order of tenths of eVs. We emphasize that this result was obtained by excluding data points where IMOM failed to converge on the ICT excitations, underlining the need for robust SCF algorithms for OO-DFT methods to be suitable for semi-automated calculations, \textit{e.g.} in screening applications. In a recent publication\cite{mazzeo2023fast}, a similar method from the $\Delta$SCF family of excited-state electronic-structure theory was applied to the ICT excitation of a Flavin molecule in a biomatrix, where one electron is transferred from a tyrosine moiety, with the mediation of an adjacent glutamine group. $\Delta$SCF methods are characterized by a low computational cost, which makes them ideal for excited-states electronic-structure calculations of complex systems like this, and the $\Delta$SCF/AMOEBA\cite{ponder2010current} QM/MM method proved capable to describe the PES of the ICT state with satisfactory results. Yet, 10 out of 30 excited state dynamics simulations failed due to SCF convergence issues. More work is going in the direction of improving the stability and reliability of OO-DFT methods\cite{schmerwitz2024saddle, levi2020variational}, with special care for cases where DFT is known to be a problematic choice, like intra-molecular CT excitations\cite{selenius2023orbital}, Rydberg excitations\cite{sigurdarson2023orbital}, conical intersections and avoided crossings\cite{schmerwitz2022variational}. With more robust SCF algorithms, it will be possible to systematically benchmark XCFs in OO-DFT against reference data, since we have shown in this study that they outperform even optimally-tuned TDA methods. Ongoing work in our lab is focused on improving the stability of the SCF convergence.

We further benchmarked the TDA and IMOM methods with the LRC-$\omega$PBE and the def2-TZVP basis for a large dataset of reference data from the literature, with similar conclusions. As discussed above, general-purpose RSH XCFs give errors on the order of 1~eV and variances of almost equal magnitude, while IMOM achieved superior results with sub-eV MSE and an MSV on the order of tenths of eV. Further testing on a subset of the data, in a pure theory-to-theory comparison, showed how optimal tuning of the $\omega$ parameter again improves the results obtained with the TDA method. However, IMOM still outperforms TDA. Finally, several RSH XCFs were benchmarked in TDA calculations with the smaller def2-SVP basis and an optimally-tuned range-separation parameter. Interestingly, a smaller basis set improves the predicted ICT excitation energies due to fortuitious error compensation, arriving at a sub-eV MSE for optimally-tuned XCFs. We conclude that IMOM is capable to provide the most accurate ICT energies among all methods tested and features systematic improvement with the basis set size. For more economic calculations we recommend optimally-tuned TDA with the $\omega$B97M-V XCF and the small def2-SVP basis set. 

\section{Acknowledgments}
The authors thank Rasma Sandra Knip\v{s}e for the contribution of several data points in Figure \ref{fig: RDA stats}.
This work is funded by the Deutsche Forschungsgemeinschaft (DFG) through the Research Training Group “Confinement Controlled Chemistry” (GRK 2376).

%\bibliography{biblio}

\providecommand{\latin}[1]{#1}
\makeatletter
\providecommand{\doi}
  {\begingroup\let\do\@makeother\dospecials
  \catcode`\{=1 \catcode`\}=2 \doi@aux}
\providecommand{\doi@aux}[1]{\endgroup\texttt{#1}}
\makeatother
\providecommand*\mcitethebibliography{\thebibliography}
\csname @ifundefined\endcsname{endmcitethebibliography}
  {\let\endmcitethebibliography\endthebibliography}{}

\begin{center}
    \includegraphics[width=\textwidth]{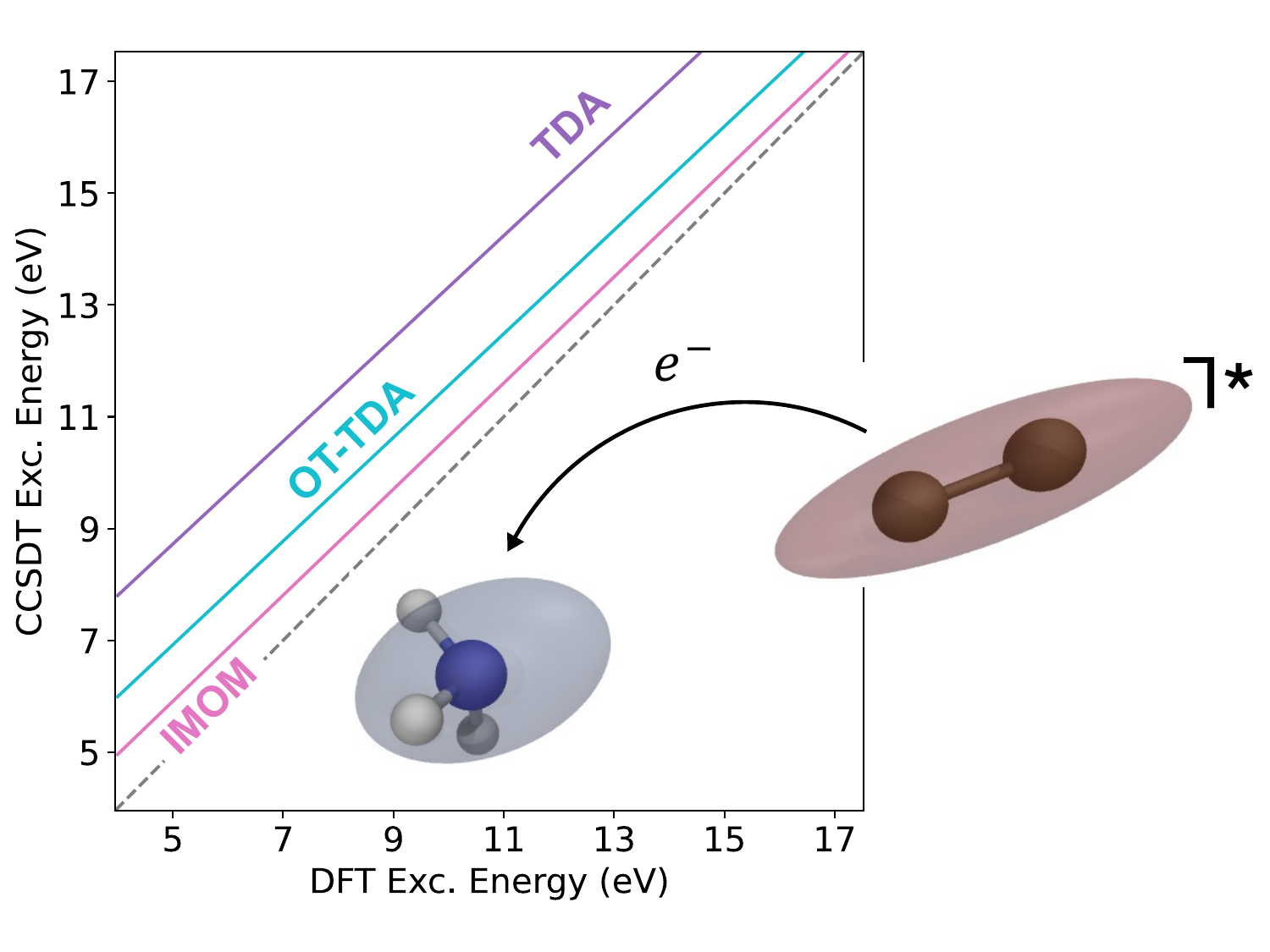}
 \label{fig: TOC}
\end{center}

\end{document}